\documentstyle[12pt]{article}
\def\beq{\begin{equation}}
\def\beqa{\begin{eqnarray}}
\def\eeq{\end{equation}}
\def\eeqa{\end{eqnarray}}
\def\lgl{\langle }
\def\nn{\nonumber }
\def\od{\odot }
\def\ra{\rightarrow }
\def\rgl{\rangle }
\def\srt{\sqrt{2}}

\input{epsf}
\begin{document}

\title{Bohmian mechanics and consistent histories}

\author{Robert B. Griffiths\thanks{Electronic mail: rgrif@cmu.edu}\\
Department of Physics,
Carnegie-Mellon University,\\
Pittsburgh, PA 15213}

\date{Version of April 28, 1999}

\maketitle

\begin{abstract}
	The interpretations of a particular quantum gedanken experiment
provided by Bohmian mechanics and consistent histories are shown to contradict
each other, both in the absence and in the presence of a measuring device.  The
consistent histories result seems closer to standard quantum mechanics, and shows
no evidence of the mysterious nonlocal influences present in the Bohmian
description.
\end{abstract}

	Bohmian mechanics \cite{bm5253,bh93,hl93,bdd95} provides a realistic
interpretation of quantum theory by means of additional ``hidden'' variables,
not part of the standard Hilbert space of wave functions, representing the
positions of particles which make up a quantum system.  All the particles have
well-defined positions at all times, and follow trajectories as a function of
time which are determined by the quantum wave function; the unitary time
development of the latter is governed by Schr\"odinger's equation.  These
particle trajectories, along with the quantum wave function, provide the basic
ontology of the theory.

\begin{figure}[t]
\epsfxsize=14truecm
\epsfbox{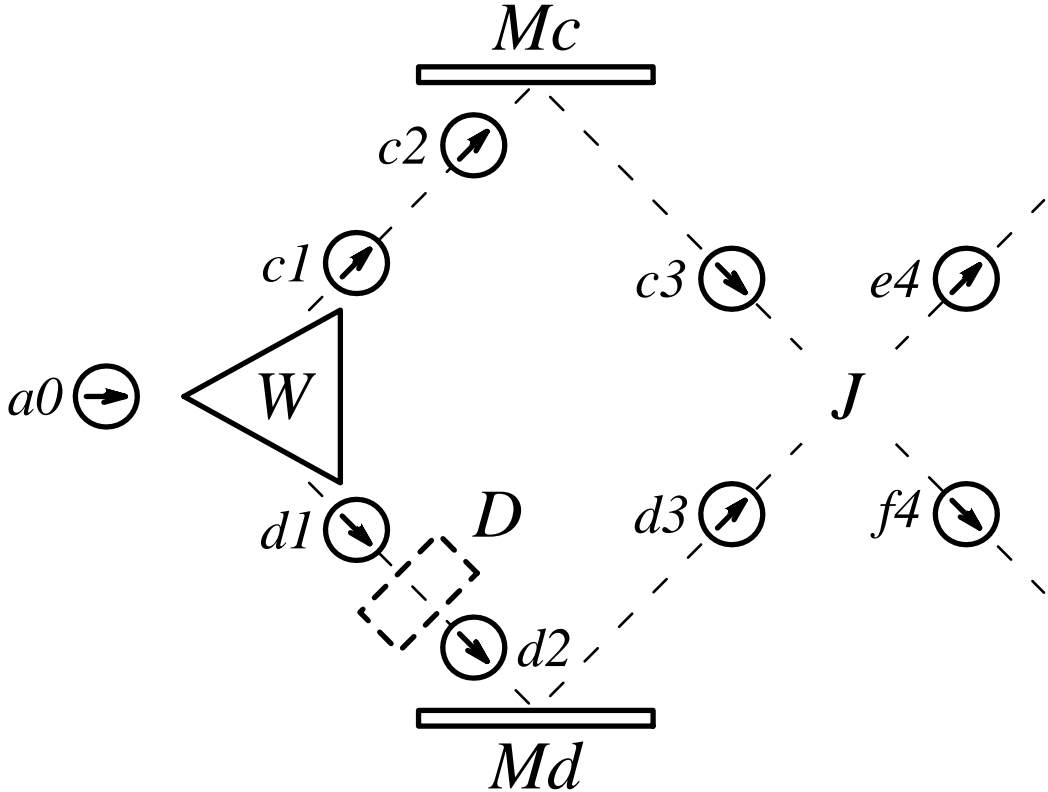}
\caption{%
Gedanken experiment in which a particle scatters from a wedge and two mirrors.
The particle wave packets at different times are shown as circles, with an
arrow indicating the direction of motion of the packet. The dashed rectangle
is a detector which may be present or absent. }
\label{f1}
\end{figure}

	Measurements do not play a fundamental role in Bohmian mechanics; they
are simply particular examples of physical processes. Hence it is quite
possible in this approach to model the measurement of a particle's position,
and ask whether the outcome of the measurement reveals the property of the
measured particle or gives a spurious result.  It was on this basis that the
physical reality of Bohmian particle trajectories was challenged by Englert et
al.\ \cite{es92}.  They asserted that the Bohm trajectory of a particle in a
bubble chamber does not necessarily coincide with the track of bubbles it
produces.  Although they did not carry out a calculation to directly support
this particular assertion, they gave detailed calculations for some other
situations in which a particular type of detector can be triggered without the
trajectory of the particle passing through it.  Their conclusion was that
Bohmian trajectory is more a metaphysical construction than an aspect of
physical reality.  See \cite{df93} for a response to this work by proponents of
Bohmian mechanics, and \cite{es93} for a reply by Englert et al.  Somewhat
later, Aharonov and Vaidman \cite{av96} made a simplified model of a bubble
chamber track and reached the same conclusion as Englert et al.  	In
addition, Dewdney, Hardy and Squires \cite{dhs93}, themselves quite sympathetic
to Bohm's approach, carried out a detailed calculation on a simple model, in
essence that shown in Fig.~\ref{f1}, and confirmed that a quantum particle can
excite a detector of this type while passing far away from it, if one accepts
the Bohmian trajectory as representing the actual particle position.  Their
article contains a defense of Bohmian mechanics as an acceptable theory despite
this somewhat surprising result.

	The consistent histories (sometimes called ``decoherent
histories'')  approach  \cite{gr84,gmh9093,om92,om94,gr96,gr98} is also a
realistic interpretation of quantum theory, but in this case the elements of
reality, or ``beables'' to use the term coined by John Bell \cite{bl87}, are
entirely quantum mechanical---wave functions or collections of wave functions
forming a subspace of the quantum Hilbert space---with no additional ``hidden''
variables.  As in Bohmian mechanics, measurements play no fundamental role in
the consistent histories approach, so it is again possible to analyze
measurements as physical processes in order to see how the measurement outcomes
are related to the properties of the measured system, and how the latter
are or are not modified by a measurement.

  	In consistent histories quantum theory, as in standard quantum
mechanics but unlike Bohmian mechanics, particles do not, in general, have
well-defined positions.  To begin with, typical wave functions are only
localized to a certain extent, and this limits the precision with which one can
ascribe a position to a particle.  In addition, even a sequence of approximate
positions, corresponding to different wave functions, as a function of time is
only permitted as a suitable quantum description when certain fairly
restrictive consistency conditions (from which this approach gets its name) are
satisfied.  Despite these limitations, consistent histories quantum theory
allows one to discuss particle positions with sufficient precision to provide a
comparison with Bohmian mechanics for the situation shown in Fig.~\ref{f1} and
discussed in \cite{dhs93}. It yields results which are directly contrary to
Bohmian mechanics, but in good agreement with those considered physically
reasonable by Englert et al., and Aharonov and Vaidman.

	The gedanken apparatus in Fig.~\ref{f1} differs slightly from that in
\cite{dhs93} in that an initial beam splitter has been replaced by a wedge $W$
in order to ensure reflection symmetry in a plane passing midway between the
mirrors $Mc$ and $Md$.  A particle initially in a wavepacket $a0$ at time $t_0$
is split into two wave packets $c1$ and $d1$ by the wedge (we ignore reflection
produced by the wedge), which are later reflected by the two mirrors so that
they cross at the position marked $J$ and eventually emerge as packets $e4$ and
$f4$ at time $t_4$.  Note that there is no physical object located at $J$; this
symbol merely indicates where it is that the two wave packets come together and
interfere with each other.  The dashed box labeled $D$ indicates the position
of a detector $D$ which is initially absent, but will be added later. The
unitary time development induced by Schr\"odinger's equation starting at $t_0$
and going on to times $t_1$, $t_2$, $t_3$, and $t_4$ takes the form
\beq
 |a0\rgl \ra (|c1\rgl + |d1\rgl)/\srt,
\label{e1}
\eeq
followed by 
\beqa
 &&|c1\rgl \ra |c2\rgl \ra |c3\rgl \ra |f4\rgl,
\nn\\
 &&|d1\rgl \ra |d2\rgl \ra |d3\rgl \ra |e4\rgl
\label{e2}
\eeqa
in an obvious notation.  One could include detectors in the output channels $e$
and $f$, but we omit them as they are not essential for the
following discussion.

	In Bohmian mechanics the particle's physical position, which is at
every time a precise point in space, as in classical mechanics, can be anywhere
inside the corresponding wave packet \cite{nt1}.  As noted by Bell \cite{bl1}
for a similar gedanken experiment, symmetry ensures that if at $t_0$ the
particle is initially above the symmetry plane inside the packet $a0$, it will
always remain above the symmetry plane.  Thus it follows wave packet $c$ as the
latter bounces from the upper mirror, along what we shall call the $c$ path,
and in the region $J$ where the $c$ and $d$ wave packets intersect, it will
again bounce and emerge in the upward-moving wave packet $e$, following the
$e$ path, rather than in the $f$ wave packet moving downward along the $f$
path.  Since there is no physical object located at $J$, many physicists will
regard this last bounce (in contrast to those which occur at the wedge and the
mirror) as counterintuitive, something Bell \cite{bl1} was aware of, but which
he ascribed to ``classical prejudice''.  In any case, it is perfectly clear in
Bohmian mechanics that if the particle emerges in the $e$ channel, it was
earlier inside the $c$ wave packet, moving along the $c$ path above the
symmetry plane.  A similar discussion applies to a particle which is
initially below the symmetry plane in $a0$: it follows the $d$ path, bounces at
$J$, and emerges in $f$.

	A consistent histories analysis can be applied to the situation shown in
Fig.~\ref{f1} by adopting a suitable family of quantum histories, as discussed
in \cite{gr96}.  It is convenient to employ a family with two histories
\beqa
 && Y_{cf} = [a0]\od[c1]\od[c2]\od[c3]\od[f4],
\nn\\
 && Y_{de} = [a0]\od[d1]\od[d2]\od[d3]\od[e4],
\label{e3}
\eeqa
of non-zero weight.  Here $[\psi]$ stands for the projection operator
$|\psi\rgl\lgl\psi|$ onto the state $|\psi\rgl$, and the sequence of projectors
separated by $\od$ symbols should be thought of in a realistic sense as events
occurring at the successive times $t_0, t_1,\ldots t_4$: the particle is in the
corresponding state, described by the corresponding wave packet, at the time in
question.  The subscripts on $Y$ are chosen to indicate the path followed by
the particle: in $Y_{cf}$ the particle follows the $c$ path up to $J$ and then
moves downward along the $f$ path. As in standard quantum mechanics, and in
contrast to Bohmian mechanics, a quantum history of this type does not ascribe
a precise position to the particle at a particular time; instead, the particle
is localized to the extent that the wave packet is localized.  While this gives
only limited precision, that is quite adequate for the present discussion: we
assume that the $|d1\rgl$ wave packet is (to an adequate approximation)
located entirely below the symmetry plane, and therefore in the history
$Y_{de}$ the particle follows the $d$ path below the symmetry plane for times
between $t_1$ and $t_3$, whereas at time $t_4$ it is described by $|e4\rgl$,
and is thus in the $e$ path.

	In addition to the histories in (\ref{e3}), it is necessary to add
other histories to make up a complete family \cite{gr96}.
There are many other histories which can be included in the family, among them:
\beqa
 && Y_{ce} = [a0]\od[c1]\od[c2]\od[c3]\od[e4],
\nn\\
 && Y_{df} = [a0]\od[d1]\od[d2]\od[d3]\od[f4].
\label{e4}
\eeqa
However, all of these additional histories carry {\it zero weight}, meaning
that they are dynamically impossible and thus do not actually occur (they occur
with zero probability).  The reason why $Y_{df}$, for example, has zero weight
is that the particle is in the wave packet $d3$ at time $t_3$, and integrating
the Schr\"odinger equation starting with $|d3\rgl$ at $t_3$ yields $|e4 \rgl$,
which is orthogonal to $|f4\rgl$, the final state of the particle in the
history $Y_{df}$.  Thus this history has zero weight because the Born rule
would ascribe zero probability to a transition from $|d3\rgl$ to $|f4\rgl$.
The weights used in consistent histories quantum theory can be thought of as
generalizations of the Born rule to histories involving more than two times;
the rule for assigning weights is explained in \cite{gr96}. 

	In order for the family to be regarded as a sample space of mutually
exclusive histories (one and only one of which actually occurs) to which
probabilities can be assigned in accordance with the laws of quantum dynamics,
it is necessary that it be a {\it consistent family} satisfying certain
mathematically precise consistency conditions, as explained in \cite{gr96}.
Since most of the histories in the family we are considering have zero weight,
the consistency conditions are easily verified.  Then, using the unitary time
transformations in (\ref{e1}) and (\ref{e2}), one can show that the only two
histories with non-zero weights, those in (\ref{e3}), both have the same weight
(again see \cite{gr96} for details). Hence, either the history $Y_{cf}$ occurs,
with probability 1/2, or $Y_{de}$ occurs, with probability 1/2.  Thus the
probability is 1/2 that the particle emerges in the $e$ path, but if it does so
it is certain (conditional probability 1) that it was earlier in the $d$ path
before it reached $J$, and not in the $c$ path, since the history $Y_{ce}$ has
zero probability.  Similarly, a particle emerging in the $f$ path (history
$Y_{cf}$) was in the $c$ path before it reached $J$, not in the $d$ path.

	Consequently, Bohmian mechanics and consistent histories, both of which
claim to go beyond standard textbook quantum mechanics in providing a {\it
realistic} description of what is happening to a quantum system as it develops
in time, in the absence of measurements, yield completely opposite and
contradictory results for the gedanken experiment in Fig.~\ref{f1}.  Bohmian
mechanics asserts that the particle will necessarily bounce when it comes to
$J$, remaining either above or below the symmetry place, thus going from $c$ to
$e$ or from $d$ to $f$.  The consistent histories approach asserts that the
particle does {\it not} bounce at $J$; instead, it follows $c$ to $f$ or $d$ to
$e$.  (Note, incidentally, that the fully quantum-mechanical treatment provided
by consistent histories yields precisely the result which Bell ascribed to
``classical prejudice''!)

	To throw further light on this contradiction, and to see how Bohmian
mechanics and consistent histories are related to textbook quantum theory, it
is useful to add a measuring device to the gedanken experiment in Fig.~\ref{f1}.
While such devices could be employed at the ends of the $e$ and $f$ channels,
this would not provide much insight into the contradiction just mentioned.  It
is more useful to add a detector at the position marked $D$ in Fig.~\ref{f1},
as this can distinguish whether the particle is traveling along path $c$ or $d$
at an intermediate time.  Of course, a measurement in which the particle simply
stopped in the detector would be of little help in discussing what happens near
$J$, so we shall assume a detector which registers the particle's passage
while perturbing its motion as little as possible. In both Bohmian mechanics
and consistent histories, the detector must be treated as a quantum mechanical
device, which means extending the Hilbert space.  In addition, for Bohmian
mechanics one must also add  additional particle positions (hidden variables).

	We take as our Hilbert space the tensor product $\cal P\otimes D$ of a
space $\cal P$ for the particle and $\cal D$ for the detector. (One could
include additional spaces for the wedge and other passive elements in
Fig.~\ref{f1}, but doing so would simply make the notation more complicated.)
Let us assume that the detector is initially in a state $|D\rgl$, and that
interaction with the particle produces a unitary time development
\beq
 |d1\rgl |D\rgl \ra |d2\rgl |D^*\rgl,
\label{e5}
\eeq
where $|D^*\rgl$ denotes a state orthogonal to $|D\rgl$ in which the detector
has detected the particle.  In all the other steps of unitary time evolution
indicated in (\ref{e1}) and (\ref{e2}), the state of the detector, whether
$|D\rgl$ or $|D^*\rgl$ does not change.  In particular, the unitary time
evolution of the initial state now takes the form
\beqa
 &&\quad \srt |a0\rgl|D\rgl \ra (|c1\rgl + |d1\rgl)|D\rgl \ra
\nn\\
 && (|c2\rgl|D\rgl + |d2\rgl|D^*\rgl) \ra 
  \cdots \ra (|f4\rgl|D\rgl + |e4\rgl|D^*\rgl).
\label{e6}
\eeqa
Note that we are making precisely the same assumption about unitary time
development as in the study \cite{dhs93} of this situation using Bohmian
mechanics; the only difference is in notation.

	Let us now ask how this situation could be given a physical
interpretation according to the ideas one finds in standard textbook quantum
theory.   One approach would be to assume that measurements are made
when the particle emerges in $e$ or $f$, following the dictum that quantum
theory only tells one the results of measurements. Based upon (\ref{e6}), one
is led to the conclusion that if the particle emerges in $e$, then it is
correlated with a detector in the state $D^*$, suggesting that the particle
triggered the detector at an earlier time, whereas if it emerges in $f$ it is
correlated with the detector state $|D\rgl$: the detector was not triggered.
This, of course, does not prove that the particle actually passed through the
detector, but one can understand how physicists, especially those who work in
the laboratory and whose results are dependent upon detectors functioning in a
certain way, might well be led to the conclusion, whether or not justified,
that a particle emerging in $e$ had earlier passed through the detector, and
hence crossed from below to above at $J$.

	But there is another, alternative, approach within the framework of
textbook quantum theory.  Since a measurement of the particle position was made
at the time $t_1$ by detector $D$, the standard von Neumann prescription would
suggest ``collapsing the wave function'' of the particle when $D$ is triggered.
That is, the detector either detects or does not detect the particle, and if it
does detect the particle, then the particle wave packet at $t_2$ should be
replaced by $|d2\rgl$.  This wave packet can then be allowed to develop in time
according to Schr\"odinger's equation, in which case it will, see (\ref{e2}),
eventually emerge in the $e$ channel: note that it does not bounce at $J$, but
instead moves from $d$ to $e$.  Granted, there is a certain embarrassment for
the standard textbook approach in that, if the detector does {\it not} detect
the particle, the wave function still collapses (which seems a bit odd), only
this time onto the wave packet $|c2\rgl$, which later emerges in path $f$,
again without anything peculiar happening at $J$.  This problem of
``interaction-free measurement'' has been known for some time within the
framework of standard quantum theory \cite{dk81}.  One can make it somewhat less
embarrassing by inserting an identical detector in the $c$ path, but we shall
not pursue the matter further, since the consistent histories approach reaches
the same conclusion as standard quantum theory without embarrassment, excuses,
or equivocation.

	Whichever approach is taken to analyzing this sort of measurement
situation from the perspective of textbook quantum theory, one arrives at the
conclusion, even if the argument is not fully rigorous, that if a measurement
by $D$ detects the particle, then it will later cross the symmetry plane near
$J$ and emerge in the $e$ channel, whereas if it is not detected, presumably
because it was in path $c$ far away from the detector, it will emerge in the
$f$ channel.  This leaves open, of course, the question of what the particle
does in the absence of detection, but seems consistent with the idea that if
one can say anything at all about what happens with the detector removed, the
particle is likely to pass through $J$ without bouncing.  Considerations of
this sort should not simply be dismissed as ``classical prejudice''; instead,
they represent thoughtful applications of a theory, textbook quantum mechanics,
whose principles are not totally consistent, but which has been successfully
used to obtain many results in agreement with laboratory experiments.

	While textbook quantum reasoning tends to be vague and somewhat
intuitive, rather than logically precise, the formulation and rules for
consistent histories quantum theory is by now just as precise and rigorous as
that of Bohmian mechanics \cite{nt2}.  And the consistent histories analysis of
Fig.~\ref{f1} with the detector in place is in complete accord with the
conclusions we have just reached on the basis of standard quantum theory.  All
one needs to do is to add appropriate detector states, $[D]$ or $[D^*]$, to the
events at the different times in the histories in (\ref{e3}), and carry out the
required analysis.  Once again, there are only two histories with non-zero
probabilities, $Y_{cf}$ with the detector state $[D]$ at all times, and
$Y_{de}$ with $[D]$ at $t_0$ and $t_1$, and the triggered detector state
$[D^*]$ at $t_2, t_3$ and $t_4$.  Each of these occurs with probability 1/2.
Consequently, the particle emerges in $e$ if and only if it was earlier in the
$d$ path and encountered and triggered the detector between $t_1$ and $t_2$.
If the particle is earlier in the $c$ path, far from the detector, it does not
trigger the detector, and it emerges in $f$.

	Note that these results with a detector present are completely
consistent with the description supplied by consistent histories when the
detector is absent.  Thus a rigorous and fully quantum-mechanical treatment
yields precisely what might have been expected from a naive classical analysis:
if a detector is placed in the $d$ path, then it detects the particle if the
particle passes along that path, and does not detect it if the particle passes
far away from it in the $c$ path, and at $J$ the particle's behavior is
precisely the same as in the absence of a detector.  The overall consistency of
the accounts with the detector present and absent is not accidental. In the the
very first paper on consistent histories \cite{gr84} it was pointed out that
the statistics associated with a {\it consistent} family can always be
confirmed, in principle, by the use of suitable idealized quantum measurements.
Of course, these measurements have to be of a sort which do not inappropriately
disturb the motion of the system being measured; (\ref{e5}) is an example.
Less gentle measurements can have other effects---but these, too, can be
analyzed using consistent histories methods.

	Now let us consider what Bohmian mechanics says about the gedanken
experiment when a detector $D$ added to the apparatus in Fig.~\ref{f1}.  Adding
such a detector breaks the symmetry, so the analysis is not as easy as before.
Indeed, it is much more complicated than either standard quantum mechanics or
the consistent histories approach, because one has to solve not only
Schr\"odinger's equation, but also also the coupled differential equations for
the (classical) trajectories of the particle whose motion is being detected
{\it and} the particle or particles which constitute the detector.  This
problem was studied in \cite{dhs93} for a detector consisting of a single
detector particle: let us call it a {\it proton} to distinguish it from the
{\it electron}, the particle whose motion through the system in Fig.~\ref{f1}
is being studied.  (The spin degrees of freedom of both particles play no
role.)

	The analysis in \cite{dhs93} required numerical integration of the
equations of motion for the (classical) positions of the proton and electron,
which are coupled to each other through the quantum wave function.  The results
obtained depend upon the initial position of the proton and that of the
electron, as well as on what happens to the detector (proton) at a later time
after the electron has moved on towards $J$.  Under certain circumstances (the
reader should consult \cite{dhs93} for details), it was found that the electron
passes through the $d$ path and triggers the detector, and then emerges in the
$e$ channel without bouncing at $J$, in agreement with what one would expect
from a consistent histories analysis and standard quantum theory.  It is also
possible for the electron to pass through the $c$ path without triggering the
detector and emerge in $f$, again without bouncing at $J$.  This, too, agrees
with the consistent histories analysis.  But this second case is a trifle odd
within the context of Bohmian mechanics since {\it had the detector been
absent} an electron passing through the $c$ path would have bounced at $J$ and
emerged in $e$, not $f$.  Thus Bohmian mechanics tells us that the later
behavior of an electron passing through the $c$ path can be influenced by the
presence or absence of the detector, despite the fact that the electron remains
at all times far away from the detector.

	Under other conditions the detector can be triggered despite the fact
that (according to Bohmian mechanics) the electron passes through the $c$ path,
and thus is never anywhere near the detector. In this case the particle will
bounce at $J$ and emerge in $e$. (For this to occur it is necessary that the
detection process does not result in a macroscopic change before the particle
has passed through the region $J$ \cite{nt3}.)  An electron emerging in the $e$
path with the detector in the triggered state is, of course, a possible outcome
at $t_4$ according to both standard quantum mechanics and consistent histories,
but then it was earlier in the $d$ path rather than in the $c$ path according
to the consistent histories analysis, and this is at least plausible if one
employs the less precise approach of standard quantum theory.  Once again, we
find the predictions of Bohmian quantum mechanics in complete disagreement with
consistent histories.

	Bohmian mechanics thus leads to a double nonlocal influence: a detector
can both affect and be affected by a particle which comes nowhere near it.  Not
only does this sort of nonlocality contradict consistent histories, it is also
in disagreement with the intuitive expectations of physicists such as Englert
and his colleagues, and Aharonov and Vaidman. Even proponents of Bohmian
mechanics, or physicists very sympathetic to this variety of quantum
interpretation, seem to find such nonlocal results a bit counterintuitive, see
the remarks in \cite{df93} (commenting on \cite{es92}) and the various comments
in \cite{dhs93}.  Their defense of Bohmian mechanics is, in essence, that that
it yields the same measurement statistics for particle positions as does
standard quantum theory, and that standard quantum theory has no sensible way
of saying what happens to a particle aside from, or prior to, a measurement.
Consequently, there is no way in which standard quantum mechanics can
contradict the assertions of Bohmian mechanics about a particle's trajectory.
Thus at the conclusion of \cite{dhs93} one finds: ``The {\it only known} way to
discuss quantum mechanics in terms of trajectories consistently is to use the
Bohm approach\dots'' [emphasis in the original].

	A basic problem with this defense is that standard quantum mechanics is
in practice not so much a set of well-defined and carefully codified principles
as it is a collection of phenomenological rules learned by students through a
process of working out examples, with solutions marked right or wrong by the
professor.  In such a discipline a significant part of the practitioner's skill
resides not in the mathematical formulas and statements of basic principles,
but rather in a sort of physical intuition which says which rules are to be
applied in which circumstances.  Even the experts have admitted that standard
quantum mechanics contains unresolved inconsistencies, and they have sometimes
discussed them in detail \cite{wg63}.  This has been one of the motivations
behind the development of Bohmian mechanics, and of consistent histories.
Consequently, agreement with standard quantum mechanics is a somewhat vague
notion, depending upon which rules one thinks are most important, etc.  As the
preceding analysis has shown, there are grounds for doubting whether Bohmian
mechanics agrees with standard quantum mechanics when a measuring device is
added to Fig.~\ref{f1}, and while these doubts can be suppressed by adopting a
sufficiently restrictive view of what standard quantum can really assert, this
then runs the danger of ignoring the physical intuition of practicing
physicists.  On the one hand, one must acknowledge that many aspects of
standard quantum theory do make its descriptions of physical phenomena a
``great smoky dragon''.  On the other hand, experiments in quantum optics are
unlikely to succeed if the experimenters do not have some understanding of the
principles involved in building and operating detectors, and this suggests that
the concerns expressed in \cite{es92} have not been adequately dealt with in
the response from proponents of Bohmian mechanics that one finds in
\cite{df93}; see also \cite{es93}.

	In any case, Bohmian mechanics is {\it not} the only consistent way to
discuss quantum trajectories, at least if one is willing to allow the sort of
(somewhat) delocalized description inherent in a quantum wave packet, such as
$c1$ in Fig.~\ref{f1}, which is still precise enough to clearly distinguish
path $c$ from path $d$. These wave packets can, as we have seen, be combined
into consistent histories to produce physically sensible results which agree
with what most quantum physicists would expect, and show no evidence for
particles bouncing in the middle of empty space, nor peculiar nonlocal
influences of particles on detectors, or detectors on particles.  Thus by using
consistent histories one reaches the same results as standard quantum theory
without employing the vague arguments and intuitive jumps for which it has been
justly criticized.  The advent of consistent histories quantum theory sets a
higher standard of quantum interpretation against which Bohmian mechanics can be
and should be measured.

	Bohmian mechanics and consistent histories are both attempts to remove
the ambiguities and contradictions inherent in textbook quantum theory by
giving a mathematically precise and physically realistic interpretation.  Both
of them are formulated in clear mathematical terms, and are free of logical
inconsistencies, so far as one knows at present.  The picture of physical
reality which they provide, on the other hand, is very different. It is hoped
that the preceding discussion will assist the reader in assessing which
approach is more satisfactory. 

	Acknowledgment. The author is indebted to O. Cohen, C. Dewdney, and S.
Goldstein for comments on the manuscript, and to them and to L. Hardy for
helpful discussions about Bohmian mechanics.  The research described here was
supported by the National Science Foundation Grant PHY 96-02084.


\begin{thebibliography}{11}
\bibitem{nt1} 
	While the particle follows a deterministic trajectory, its actual
position is not known.  If this ignorance is expressed as a probability, the
probability density for the position at a particular time is proportional to
the absolute square of the quantum wave function, at that time, as a function
of position.
\bibitem{nt2}
	From time to time the claim is made that consistent histories quantum
theory is inconsistent, or at least logically suspect.  See \cite{gr98} for
references to a number of criticisms of this sort, and a response from a
consistent histories perspective.  It is important in this connection not to
overlook the ``one family'' (``one framework'', ``one logic'') rule, for which
see p.~163 of \cite{om94}, Sec.~V of \cite{gr96}, and \cite{gr98}.
\bibitem{nt3}
	As explained in \cite{es92}, there is no problem in principle in
constructing a microscopic detector which records the passage of a particle and
whose state is read out at a later time as a macroscopic signal.  The
triggering of a detector of this type, with a delayed recording of the result,
would thus normally be interpreted as experimental evidence that a particle
passed through the detector, whether or not the transformation to a
macroscopic signal occurs before the particle reaches $J$.  In the consistent
histories approach one can describe the detection event as occurring as (or
just after) the particle passes through the microscopic detector, and, in
contrast to Bohmian mechanics, the actual time at which microscopic information
is transformed into a macroscopic signal does not make any difference. If the
apparatus has been properly constructed, the earlier microscopic state can be
inferred from the later macroscopic signal, and detection in either sense
occurs if and only if the particle passes through the detector.  (For some
additional discussion of measurement processes using consistent histories, see
\cite{gr96,gr98}.)
\end{thebibliography}
\end{document}